\documentclass[twocolumn,floats,superscriptaddress,PRX]{revtex4-2}

\usepackage[pdftex]{graphicx}
\usepackage{amsmath}
\allowdisplaybreaks
\usepackage{amsfonts}
\usepackage{mathrsfs}
\usepackage[utf8]{inputenc}
\usepackage{color}
\usepackage{xcolor}
\usepackage[normalem]{ulem}
\usepackage{upgreek}

\begin{document}

\title{Chirality, anisotropic viscosity and elastic anisotropy in three-dimensional active nematic turbulence}

\author{Nika Kralj}
\affiliation{Faculty of Mathematics and Physics, University of Ljubljana, Ljubljana, Slovenia}
\author{Miha Ravnik}
\affiliation{Faculty of Mathematics and Physics, University of Ljubljana, Ljubljana, Slovenia}
\affiliation{
Department of Condensed Matter Physics, Jožef Stefan Institute, Ljubljana, Slovenia
}
\author{Žiga Kos}
\email{ziga.kos@fmf.uni-lj.si}
\affiliation{Faculty of Mathematics and Physics, University of Ljubljana, Ljubljana, Slovenia}
\affiliation{
Department of Condensed Matter Physics, Jožef Stefan Institute, Ljubljana, Slovenia
}
\affiliation{International Institute for Sustainability with Knotted Chiral Meta Matter (WPI-SKCM$^2$), Hiroshima University, Higashi-Hiroshima, Japan}

\date{\today}

\begin{abstract}
\vspace{0.5 cm}
\section*{ABSTRACT}
Various active materials exhibit strong spatio-temporal variability of their orientational order known as active turbulence, characterised by irregular and chaotic motion of topological defects,  including colloidal suspensions, biofilaments, and bacterial colonies.In particular in three dimensions, it has not yet been explored how active turbulence responds to changes in material parameters and chirality.Here, we present a numerical study of three-dimensional (3D) active nematic turbulence, examining the influence of main material constants:  (i) the flow-alignment viscosity, (ii) the magnitude and anisotropy of elastic deformation modes (elastic constants), and (iii) the chirality. Specifically, this main parameter space covers contractile or extensile, flow-aligning or flow tumbling, chiral or achiral elastically anisotropic active nematic fluids. The results are presented using time- and space-averaged fields of defect density and mean square velocity. The results also discuss defect density and mean square velocity as possible effective order parameters in chiral active nematics, distinguishing two chiral nematic states---active nematic blue phase and chiral active turbulence. This research contributes to the understanding of active turbulence, providing a numerical main phase space parameter sweep to help guide future experimental design and use of active materials.
\end{abstract}

\maketitle

\section*{INTRODUCTION}
Active fluids are diverse synthetic and biological materials ~\cite{shankar2022topological, SanchezT_Nature491_2012, WensinkHH_ProcNatlAcadSci109_2012,HardouinJ_SoftMatter16_2020,wittmann2023collective,peng2016command,sokolov2007concentration,dombrowski2004self,kokot2017active,karani2019tuning}. Active nematics are characterized by orientational order along the director vector $\mathbf{n}$ and propelled by anisotropic active stress~\cite{HatwalneY_PhysRevLett92_2004,VoituriezR_EurophysLett70_2005}. A notable and rather ubiquitous dynamic state exhibited by active nematics is active turbulence~\cite{AlertR_AnnuRevCondensMatterPhys13_2022}, characterized by continuous spatio-temporal defect proliferation and annihilation. This phenomenon is governed by the intricate interplay between defects within the director field and the structures of the velocity field~\cite{head2024spontaneous}. In three dimensions, the structure of active turbulence comprises a dynamic rewiring network of defect lines and loops~\cite{DuclosG_Science367_2020,UrzayJ_JFluidMech822_2017,KrajnikZ_SoftMatter16_2020,singh2023numerical}, where the diverse range of possible defect shapes and their interplay with the velocity field pose significant challenges for its understanding, control and possible applicability.

Approaches for studying three-dimensional (3D) active turbulence include theoretical models of individual defect loops~\cite{BinyshJ_PhysRevLett124_2020,LongC_SoftMatter17_2021,houston2022defect}, spectral analysis~\cite{UrzayJ_JFluidMech822_2017,KrajnikZ_SoftMatter16_2020}, and defect tracking and extracting the statistics of linked loops~\cite{romeo2023vortex}, defect curvature and length~\cite{digregorio2024coexistence,BinyshJ_PhysRevLett124_2020}.
Another possible approach is to extract the mean-field observables such as average defect density and squared velocity. The scaling of such observables with the activity was determined for single elastic constant and constant viscosity parameters~\cite{kralj2023defect,digregorio2024coexistence}. However, a systematic study of 3D active nematic turbulence for different elasticity, viscosity, and chirality material parameters also at different values of activities has not been performed yet. The role of chirality in active turbulence is also relevant as the active matter materials are often weakly chiral~\cite{furthauer2012active,whitfield2017hydrodynamic}.
As shown also for 2D active nematics~\cite{giomi2015geometry,AlertR_NaturePhys16_2020,thampi2014vorticity}, a systematic numerical study can lead to better experimental control of active mater  with different material properties and serve as a benchmark for theoretical models of activity-dependent dynamic regimes, instabilities, and coupled structures in the flow and orientation field.

In this paper, we show three-dimensional active nematic turbulence at different values of the main material parameters -- i.e. (i) the flow-alignment viscosity parameter, (ii) the elastic constants of splay, twist, and bend deformation modes, and (iii) the intrinsic chirality. Specifically, chirality is introduced via chiral elastic energy contribution. From numerical simulations of active dynamics, defect density and mean square velocity are extracted. The simulations are performed for contractile and extensile active materials and show distinct scalings with alignment parameter and average elastic constant, while elastic anisotropy has little effect on the mean-field averaged observables of active turbulence. For chiral active nematic turbulence, we show that increasing inverse chiral pitch $q_0$ increases the defect density and decreases the mean square velocity. At high values of $q_0$, a transition between the active nematic blue phase and chirality-affected active turbulence is observed. Distinctly, in the active nematic blue phase we observe effective jamming of the defect lattice and a strong drop in the magnitude of the flow field.

\section*{RESULTS AND DISCUSSION}
\begin{figure*}[ht!]
  \includegraphics[width=\textwidth]{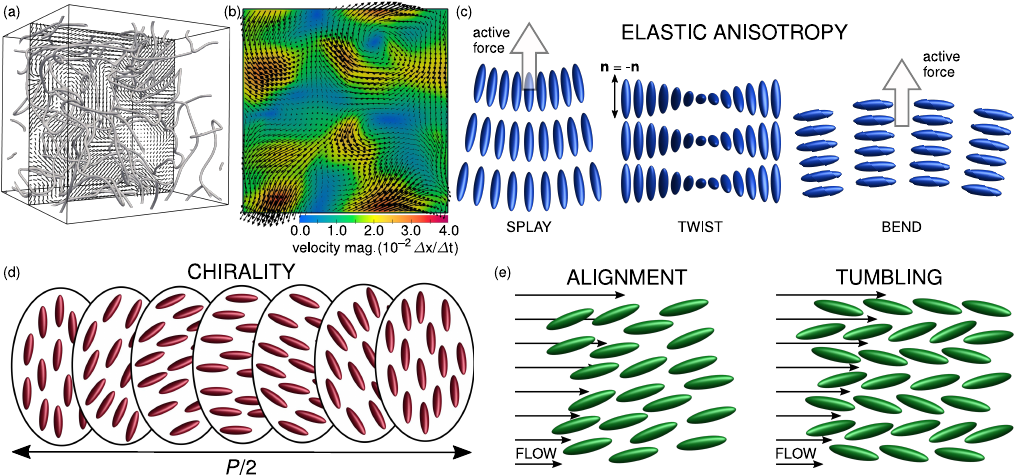}
  \caption{Active nematic turbulence in a three dimensional system. (a,b) A snapshot of an active turbulence showing (a) a disclination network (gray isosurfaces) with a director field cross-section (black rods) and (b) a cross section of the velocity field. (c--e) A schematic representation of mechanisms of (chiral) active turbulence: (c) different elastic modes of director deformations--splay, twist and bend--with the direction of the active force generated by the splay and bend distortions, (d) chiral structure of the director field with pitch length $P$, and (e) flow alignment and flow tumbling regime of the director field in shear flow.} 
  \label{fig:fig1}
\end{figure*}

We performed numerical simulations of 3D active nematic turbulence using the Beris-Edwards model of nematodynamics and the active stress tensor (Methods). We numerically solve the Q-tensor and the velocity field dynamics on a periodic grid using finite difference and lattice-Boltzmann numerical approaches, respectively\cite{CarenzaLN_ProcNatlAcadSci116_2019, ZhangR_NatCommun7_2016,thampi2014vorticity,kralj2023defect,KrajnikZ_SoftMatter16_2020,CoparS_PhysRevX9_2019}. The main mechanisms of active turbulence that we focus on are visualised in Fig.~1.

Q-tensor field is shown in Fig.~1(a), where black rods show the director field as the main eigenvector of the Q-tensor and the gray isosurfaces show the degree of order representing the main eigenvalue. The degree of order is reduced near the core of disclination lines, which are string-like disordered structures spanning the simulation box. Figure~1(b) shows the velocity field and its magnitude for a selected snapshot of  active turbulence. The nematic alignment has three main deformation modes, splay $\left(\nabla\cdot\mathbf{n}\right)^2$, twist $\left[\mathbf{n}\cdot\left(\nabla\times\mathbf{n}\right)-\frac{2\pi}{p_0}\right]^2$, and bend $\left[\mathbf{n}\times\left(\nabla\times\mathbf{n}\right)\right]^2$, each with its own elastic constant $K_1$, $K_2$, and $K_3$, respectively. Twist mode can be spontaneously favoured in chiral nematic fluids by incorporating a finite intrinsic chiral pitch length $p_0$ (Fig.~1(d)).
Active force is computed as $-\zeta\nabla\cdot Q$ and is generated by the splay and bend deformations of the active constituents alignment~\cite{whitfield2017hydrodynamic}. The force direction in Fig.~1(c) is shown for extensile active materials ($\zeta>0$), opposite direction is expected for contractile materials ($\zeta<0$). Anisotropic viscosity of nematic fluids is in the Beris-Edwards model related to the flow-aligning parameter $\chi$, which is typically dependent on the shape of nematic building blocks and describes if they are aligning in flow gradient at a fixed angle, or constantly tumbling (Fig.~1(e)). Note that in principle, a general (passive or active) nematic fluid has 6 viscosity coefficients, 5 of which are independent~\cite{deGennesPG_1993}. While the concepts of elastic anistropy, chirality, and flow alignment are well understood in equilibrium and driven nematic fluids, their effects on the irregular state of active turbulence is less understood, particularly in three dimensions. Here, we show how the properties of active turbulence depend on the material coupling constants determining the strength or anisotropy of elasticity, viscosity, and chirality.

\subsection*{Extensile, contractile, aligning, and tumbling nematics}

\begin{figure*}[ht!]
  \includegraphics[width=\textwidth]{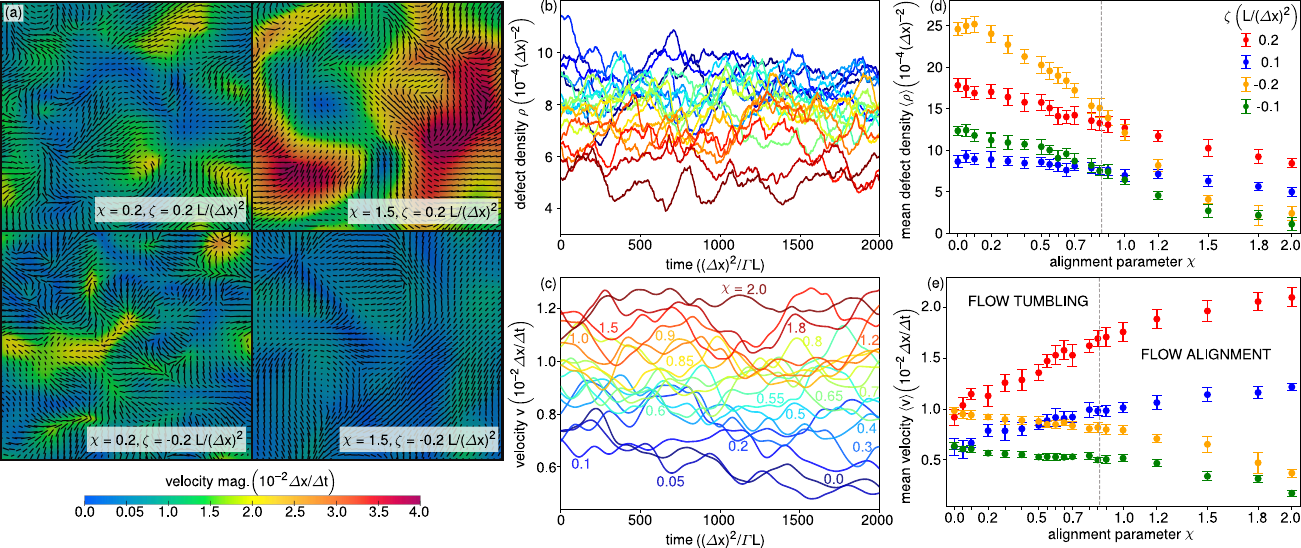}
  \caption{Dynamics of three-dimensional active nematic turbulence for extensile and contractile materials at different aligning parameters. (a) Cross-section of the velocity field magnitude (color map) and the director field (black rods) at two different activities $(\zeta = 0.2 \, L/(\Delta x)^2 \text{ and }\zeta = -0.2 \, L/(\Delta x)^2)$ and two different alignment parameters $(\chi=0.2$ \text{ and } $\chi=1.5)$. (b,c) Defect density and volume-averaged root mean square velocity over time in a dynamic steady state for activity $\zeta = 0.1 \, L/(\Delta  x)^2$ and different alignment parameters, with the lowest plotted value of $\chi = 0.0$ in dark blue colour and the highest value of $\chi=2.0$ in the dark red colour. (d,e) The dependence of the mean defect density and the mean velocity on the alignment parameter $\chi$. For $\zeta=0.1(\Delta x)^2$, the mean defect density and the mean velocity are directly computed from panels (b) and (c), respectively. The mean defect density decreases with the alignment parameter for extensile and contractile materials. The mean velocity shows an increasing trend with the alignment parameter for extensile materials ($\zeta>0$) and a decreasing trend for contractile materials ($\zeta<0$). Error bars in (d,e) represent the standard deviation originated by the time-averaged defect density and root mean square velocity, respectively.
  }
  \label{fig:alignment}
\end{figure*}

The defect density and the average flow magnitude in active turbulence are affected by the viscosity parameters of the nematic fluid. Here, we vary the flow-aligning parameter $\chi$, effectively modelling the flow-tumbling ($\chi<\chi_\text{t}$) and flow-aligning ($\chi>\chi_\text{t}$) nematic fluids~\cite{deGennesPG_1993}, where the transition between flow-aligning and flow-tumbling regime in our simulations corresponds to $\chi_\text{t} = \frac{9S_\text{eq}}{4+3S_\text{eq}}=0.86$. At the highest value of the alignment parameter in the simulation ($\chi=1.5$), the cross-section of the director field and the velocity magnitude field show deformation at a larger scale compared to the lowest value of the aligning parameter at $\chi=0.2$ (Fig. \ref{fig:alignment}(a)). During the simulation, both the defect density (Fig. \ref{fig:alignment}(b)) and the mean square velocity (Fig. \ref{fig:alignment}(c)) fluctuate in time due to finite size effect of the simulation box. We consider the effect of the alignment parameter both for the extensile $(\zeta > 0)$ and the contractile $(\zeta < 0)$ active nematics. Independently from the sign of the active stress, the mean defect density is decreasing with an increase of the alignment parameter $\chi$ (Fig. \ref{fig:alignment}(d)). Differently, we observe that the mean velocity decreases with the alignment parameter for contractile active nematics and increases for extensile active nematics (Fig. \ref{fig:alignment}(e)). Different values of the flow-aligning parameter result in different Ericksen-Leslie coefficients of the anisotropic viscosity tensor (see Methods).
Similarly to 2D active nematics with substrate friction~\cite{thijssen2020active}, some change of active behaviour is observed directly at the transition between flow-aligning and flow-tumbling regime at $\chi_\text{t} =0.86$. Mean velocity dependence on the alignment parameter $\chi$ in Fig.~\ref{fig:alignment}(e) gradually changes its slope around $\chi_\text{t}$.

\subsection*{Role of elastic constants in active turbulence}
Simulations at different values of the nematic elastic constants show two main results:
(i) the defect density scales approximately linearly with the average elastic constant, and (ii) changing the elastic anisotropy---different elastic penalties of the splay, twist, and bend director deformation modes---has a minor effect on the active nematic defect density and average velocity magnitude.

\begin{figure}[h!]
  \includegraphics[width=8cm]{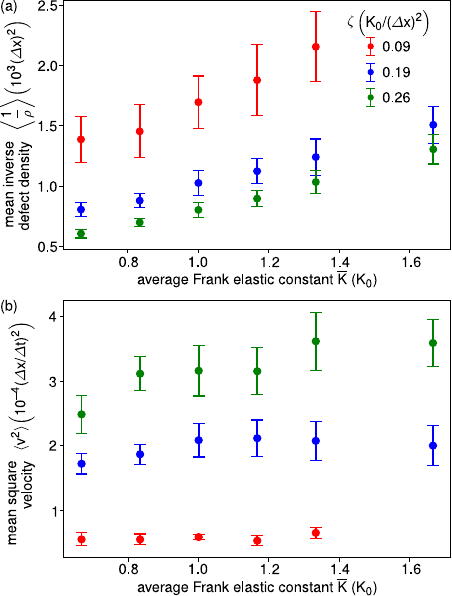}
  \caption{Defect density and mean square velocity as dependent on the average elastic constant. (a) Steady state mean inverse defect density and (b) steady state mean inverse velocity squared, both for three different activities. The average elastic constant is expressed in units of $K_0$ and is computed for different twist vs splay and bend elastic constants (see Methods). Error bars represent the standard deviation originated by the time-averaged quantities.
  }
  \label{fig:elastic_constants_L21}
\end{figure}

Figure 3 shows the defect density and mean square velocity when varying the magnitude of the average elastic constant. We observe that in 3D active turbulence the mean inverse defect density increases with increasing average Frank elastic constant $\overline{K}$ (Fig.~\ref{fig:elastic_constants_L21}(a)),
in agreement with the scaling of $\rho\sim\frac{|\zeta|\gamma_1}{\eta \overline{K}}$,
predicted by considering the role of defect density on the line tension, drag force, and defect self-propulsion~\cite{kralj2023defect}, where $\gamma_1$ is the rotational viscosity and $\eta$ the effective isotropic viscosity.
Likewise, the mean square velocity (Fig.~\ref{fig:elastic_constants_L21}(b)) also shows a slowly increasing trend with the average elastic constant, in line with the scaling
$\langle v^2\rangle\propto|\zeta|\overline{K}$, obtained by considering the self-propulsion velocity~\cite{BinyshJ_PhysRevLett124_2020,GiomiL_PhilTransRSocA372_2014} in the tubular neighborhood  of a disclination with a radial size of $\sim1/\sqrt{\rho}$.
The observed behaviour is  in agreement with results from \emph{two-dimensional} active turbulence,
where the inverse defect density and the mean square velocity are also reported to be proportional to the magnitude of the elastic constant \cite{thampi2014vorticity,hemingway2016correlation}.

\begin{figure*}[ht!]
  \includegraphics[width=\textwidth]{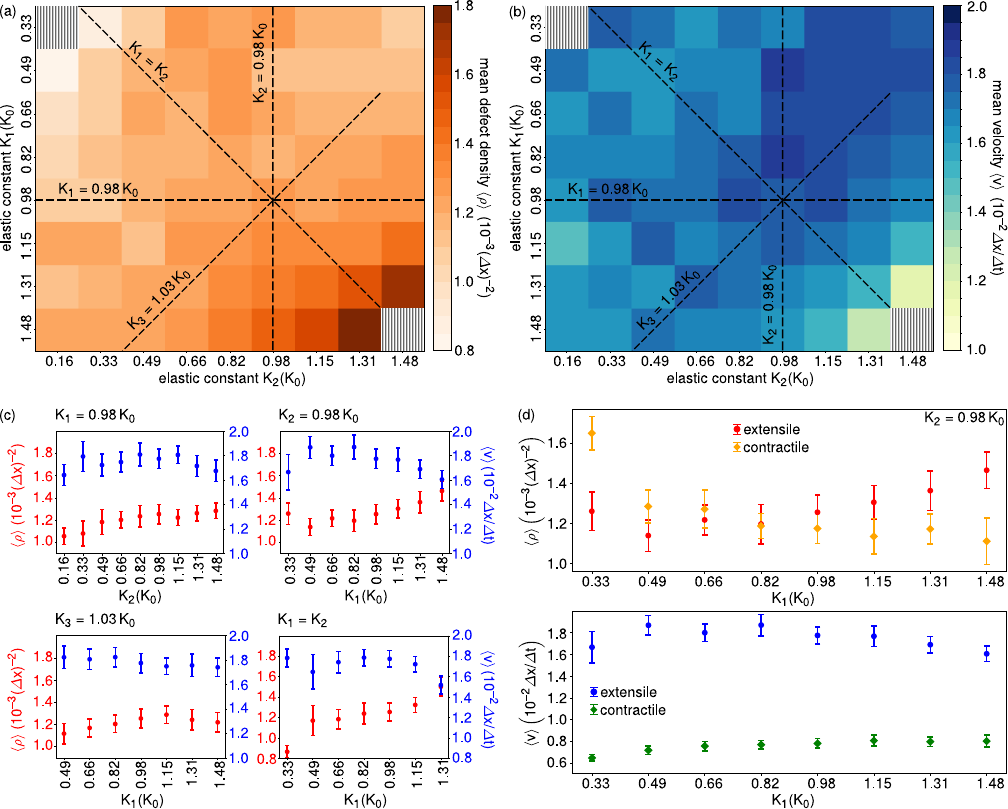}
  \caption{Effect of elastic anisotropy--different elastic constants--on three-dimensional active turbulence. (a) Mean defect density, (b) root mean square velocity and (c) selected cross-sections of the heat maps in (a,b), for the same average Frank elastic constant $\overline{K}= K_0$ and of fixed activity $\zeta = 0.2 \, L/(\Delta x)^2$. The data is represented using Frank elastic constants $K_1$ and $K_2$ (note that constant $\overline{K}$ implies $K_3= 3\overline{K}-K_1 - K_2$). Small variability is observed, which is comparable to the fluctuations of the defect density and the mean velocity in time, as inherent to the chaotic active turbulent dynamics in finite volume. (d) Mean defect density and mean velocity of extensile $(\zeta = 0.2\, L/(\Delta x)^2 )$ and contractile $(\zeta = -0.2\, L/(\Delta x)^2 )$ systems for constant twist mode $(K_2)$. 
  Error bars in (c,d) represent the standard deviation originated by the time-averaged quantities.
  }
  \label{fig:elastic_constants_K123}
\end{figure*}

Figure~\ref{fig:elastic_constants_K123} shows the role of elastic anisotropy between the splay ($K_1$), twist ($K_2$) and bend ($K_3$) modes. We change the three elastic constants under the assumption of a fixed average elastic constant $\overline{K}=(K_1+K_2+K_3)/3$ (see also Methods). Different values of the elastic constants provide relatively comparable results as seen on colormaps for the mean defect density in Fig.~\ref{fig:elastic_constants_K123}(a) and mean velocity in Fig.~\ref{fig:elastic_constants_K123}(b). The results in Fig.~\ref{fig:elastic_constants_K123}(c) show that the elastic anisotropy in the considered range has little effect on the dynamics of the active turbulence since both mean defect density and mean velocity stay roughly equal under the condition of same $\overline{K}$. With the condition of equal splay and twist modes ($K_1=K_2$, in Fig.~\ref{fig:elastic_constants_K123}(c), a slight increasing trend  of the defect density is observed with increasing $K_1$ and $K_2$ and decreasing $K_3$. 
Fig.~\ref{fig:elastic_constants_K123}(d) shows the role of anisotropy between splay and bend elastic constants for extensile and contractile systems.
For extensile systems, a weakly increasing trend of mean defect density with increasing splay elastic constant can be observed. Contrary, for contractile active nematics, defect density gradually decreases with increasing $K_1$ at constant average Frank elastic constant.
The main difference between extensile and contractile nematics is that the mean velocity is approximately 2-times larger in extensile systems for the same absolute value of activity. Similar behaviour was observed also in 2D active nematics~\cite{thampi2014vorticity}.

\subsection*{Chiral active turbulence and transition to active blue phase}

\begin{figure*}[p!]
  \includegraphics[width=\textwidth]{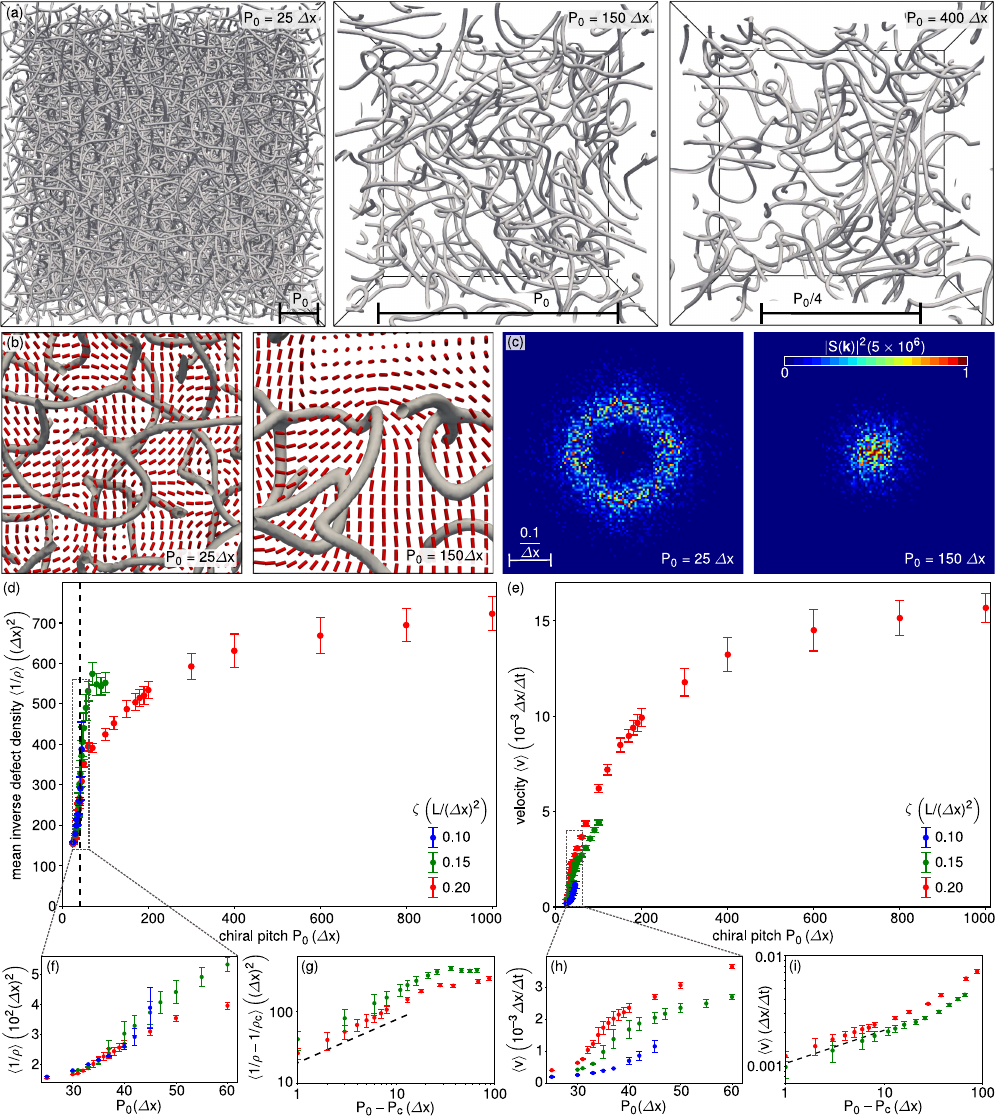}
  \caption{Chiral active turbulence and active blue phase. (a) Defect network at three different values of chiral pitch $P_0$.
  (b) Zoom of defect network at two different values of chiral pitch $P_0$, with cross-section of director field in red. 
  (c) Structure factor $|S(\mathbf{k})|^2$ at two different values of chiral pitch $P_0$, on cuts along $k_y = 0$. The structure factor was calculated from the Fourier transformation of the highest eigenvalue of the Q-tensor.
  (d) Mean inverse defect density dependence on the chiral pitch $P_0$ for three different activities. 
  Dashed line shows the value of the transition in a passive nematic at $P_\text{c}\approx 40\,\Delta x$.
  (e) Root mean square velocity dependence on the chiral pitch $P_0$. (f, h) A zoomed-in area of mean inverse defect density and root mean square velocity from $P_0 = 25\,\Delta x$ to $P_0=60\,\Delta x$. (g, i) Log-log plot of mean inverse defect density and root mean square velocity, respectively. The data in the log-log plots is plotted relative to the critical pitch length $P_\text{c}$ and the critical defect density. The values of $P_\text{c}=32\,\Delta x$ and $1/\rho_\text{c}=160\,\Delta x$ were chosen for the activity $\zeta=0.2\, L/(\Delta x)^2$, and the values of $P_\text{c}=34\,\Delta x$ and $1/\rho_\text{c}=170\,\Delta x$ for the activity $\zeta=0.15\, L/(\Delta x)^2$. 
  The dashed line in (g) has a slope of $ x^{0.6}$ and the dashed line in (i) a slope of $x^{0.3}$. 
  Error bars in (d-i) represent the standard deviation originated by the time-averaged quantities.
  }
  \label{fig:q0_rho_error}
\end{figure*}

Material chirality in nematic systems can emerge as a result of chiral nematic building blocks, chiral dopants, or--in active systems--as a results of chiral dynamics of the active agents~\cite{furthauer2012active,whitfield2017hydrodynamic}. In bulk 3D chiral active nematic, the intrinsic chirality increases the defect density and reduces the mean square velocity of active turbulence, notably already in the low chirality (i.e. large pitch) regime, as shown in Fig.~\ref{fig:q0_rho_error}. For example, for the chiral pitch equal to $p_0=200\,\Delta x$ (i.e. 89-times the active length $l_\text{a}=\sqrt{L/\zeta}$) at activity $\zeta=0.2\,L/(\Delta x)^2$, the defect density is increased by $48\%$ and the root mean square velocity decreased by $42\%$ compared to the dynamic steady state value for achiral active turbulence (Fig.~\ref{fig:alignment}(d,e)). 
The effect of intrinsic chirality on defect density and mean square velocity is larger than the variation of $\sim 30\,\%$ that was observed for elastic anisotropy in Fig.~\ref{fig:elastic_constants_K123}. One possible explanation is that for elastic anisotropy in Fig.~\ref{fig:elastic_constants_K123}, the elastic deformation modes are energetically unfavorable; however, for intrinsic chirality in Fig.~\ref{fig:q0_rho_error} nematic twist distortions are energetically preferred, which can have a greater effect on the emerged structure.

Upon further increasing chirality (i.e. reducing pitch) the passive chiral nematic is known to transition into 3D chiral orientational structures known as chiral blue phases I, II and III, and we observe a similar transition in active chiral nematics. Increasing chirality at fixed activity causes a steady increase of the defect density (Fig.~\ref{fig:q0_rho_error}(d,f,g)), up to a structure where defect lines of the active turbulence jam into a defect network, that one could identify as an effective \emph{active} blue phase III (Fig.~\ref{fig:q0_rho_error}(a), first panel). 
Similarly as for the passive blue phase III~\cite{henrich2011structure}, structural factor shows that the disclination network in active blue phase has no crystal-like order (Fig.~\ref{fig:q0_rho_error}(c)).
Specifically, in active blue phase at pitch $p_0=25\,\Delta x$ and $\zeta=0.2\,L/(\Delta x)^2$,  the defect density is 5-times larger than in achiral nematic turbulence case (Fig.~\ref{fig:alignment}(d)). An even stronger indication of an effective structural transition is the root mean square velocity (Fig.~\ref{fig:q0_rho_error}(e,h,i)) that drops to $2\%$ compared to the achiral case at $\zeta=0.2\,L/(\Delta x)^2$. 
For the numerical material parameters of our simulation, the transition to a passive blue phase is observed at a critical pitch length $P_\text{c}\approx40\,\Delta x$. Activity reduces this critical pitch length to $P_\text{c}=34\,\Delta x$ at $\zeta=0.15\,L/(\Delta x)^2$ and even further to $P_\text{c}=32\,\Delta x$ at $\zeta=0.2\,L/(\Delta x)^2$. How $P_\text{c}$ is obtained from the data is explained in Methods.

The mean square velocity drop-off and effective jamming of the disclination network can be explained by the difference of the structural features of the disclination network between active blue phase and active turbulence, as shown in Fig.~\ref{fig:q0_rho_error}(b).
The local director crosssection profile of the disclination lines in the active blue phase are close to the $-1/2$ winding number, for which the active self-propulsion is well known to be zero~\cite{BinyshJ_PhysRevLett124_2020}. Selected double-twist cylinders are also shown in Fig.~\ref{fig:q0_rho_error}(b), which also do not generate a self-propulsion flow~\cite{metselaar2019topological}.
Nodal points, where 4 disclination segments meet (Fig.~\ref{fig:q0_rho_error}(b)), are further characteristic features of the blue phase III~\cite{henrich2011structure}. Dynamically, we observe that in the active blue phase at $P_0=25\,\Delta x$, the nodal points are stable for up to $\sim150\,(\Delta x)^2/\Gamma L$. Contrary, in the active chiral turbulence, nodal points appear during reconfiguration events between disclination lines and are much shorter lived -- at $P_0=150\,\Delta x$ each nodal point disappears on average after approximately $\sim13\,(\Delta x)^2/\Gamma L$.

We observe the distinct scaling of the disclination density and the mean square velocity near the transition between the active blue phase and the active turbulence. In Figure~\ref{fig:q0_rho_error}(d,e), we plot the inverse defect density and root mean square velocity dependence on the pitch length in proximity of the critical defect density $\rho_c$, critical mean velocity $v_c$ and critical chiral pitch $P_c$ of the blue phase-active turbulence transition. The slope of the graph shows that inverse defect density scales roughly as $(P_0-P_c)^{0.6}$ and mean velocity as $(P_0-P_c)^{0.3}$. Similar scaling is obtained for two different activities with a notable difference that at higher activity the blue phase-active turbulence transition occurs at smaller pitch lengths.

\section*{CONCLUSIONS}

We explore dynamic reconfiguring network of disclination lines known as active turbulence using  numerical simulations for selected main material parameters: chirality, flow alignement (anistropic viscosity), and elastic anistropy (different nematic elastic constants). The difference between extensile and contractile systems is observed in increasing or decreasing dependence of mean square velocity with the alignment parameter, showing the importance of the  shape of active nematic building blocks~\cite{brand1982theory}.
We confirm that defect density and mean square velocity are approximately inversely proportional to the magnitude of the average elastic constant, whereas the elastic anisotropy has only a small effect on the defect density and the mean square velocity of active turbulence. Though, we speculate that elastic anisotropy could affect the local structure of the defect lines ($+1/2$, $-1/2$ and twisted~\cite{BinyshJ_PhysRevLett124_2020}) in the defect network. 
As the elastic instability is known to drive structural reconfigurations in passive liquid crystals~\cite{lavrentovich2024splay}, we expect that for active nematics elastic instability might have a greater effect in confined systems at the onset of active turbulence.
While current experiments on 3D active turbulence explore bulk behaviour~\cite{DuclosG_Science367_2020}, the role of elastic anisotropy and alignment parameter is relevant also for possible future results in confined systems.

We performed simulations of an active chiral nematic and show the effective structural transition between chiral active turbulence and the active blue phase. The structures are distinct from each other and are separated by a continuous structural phase transition that we characterize by measuring the average inverse defect density and mean square velocity. Beyond the results reported here, the observed active blue phase dynamics would be (i) interesting to compare with blue phase dynamics due to thermal fluctuations~\cite{pivsljar2022blue} and (ii) explored in the context of driving with external field,  such as electric or magnetic fields~\cite{kikuchi2004fast} or activity gradients ~\cite{shankar2024design}. An interesting aspect is also the difference between the transition to active turbulence in blue phases and in modulated cholesteric phase, for which a linear hydrodynamic instability was predicted for extensile materials~\cite{whitfield2017hydrodynamic}.
While experimentally engineering 3D active blue phase materials can lead to very novel materials and phenomena, the effects of weak chirality that we show in the paper might be important also for present active nematic materials, since building blocks and processes in active matter are often chiral~\cite{furthauer2012active,whitfield2017hydrodynamic},
which could be even further amplified by introduction of chiral dopants. Additionally, chiral symmetry breaking allows for additional active stresses~\cite{kole2021layered,maitra2020chiral,hoffmann2020chiral,markovich2019chiral}. 
An open problem for future research is also the possibility of an active blue phase with symmetries of blue phase I or blue phase II~\cite{yamashita2022structure}.

More generally, this work is a contribution towards understanding the material-dependence of active nematic turbulence, to aid the experimental design and theoretical advances of active nematic phase~\cite{skogvoll2023unified,pratley2023three,velez2024probing}.

\section*{METHODS}

\subsection*{Model equations of active nematodynamics}
We simulate mesoscopic continuum description of active nematics using the adapted Beris-Edwards approach for active nematodynamics ~\cite{DoostmohammadiA_NatCommun9_2018, CoparS_PhysRevX9_2019, CarenzaLN_ProcNatlAcadSci116_2019}. 
The nematic order is described by a traceless tensor order parameter $Q_{ij}$, with the director $\mathbf{n}$ as the main eigenvector, and evolves as 
\begin{equation}
    \label{eq:q_dynamics}
    \left(\partial_t + v_k \partial_k\right)Q_{ij} - S_{ij} = \Gamma H _{ij},
\end{equation}
where $\mathbf{v}$ is the fluid velocity and $\Gamma$ is the rotational viscosity coefficient. The generalized advection term $S_{ij}$ couples the nematic order and fluid velocity
\begin{align*}
    S_{ij} =& \left(\chi D_{ik} -\Omega_{ik} \right)\left(Q_{kj} + \frac{\delta_{kj}}{3} \right) \\
    &+ \left(Q_{ik} + \frac{\delta_{ik}}{3} \right)\left(\chi D_{kj} +\Omega_{kj} \right) \\
    &- 2\chi \left(Q_{ij} + \frac{\delta_{ij}}{3}\right) Q_{kl} W_{lk} ,
\end{align*}
where $D_{ij}$ and $\Omega_{ij}$ are symmetric and antisymmetric part of the velocity gradient tensor $W_{ij} = \partial_i v_j$, and $\chi$ is the alignment parameter. The molecular field $H_{ij}$ drives system towards equilibrium of $Q_{ij}$ 
$$H_{ij} = -\frac{\delta F }{\delta Q_{ij}} + \frac{\delta_{ij}}{3} \mathrm{Tr}\frac{\delta F }{\delta Q_{ij}},$$
where $F$ is the Landau-de Gennes free energy
\begin{align*}
    F &= \int \left( \frac{A}{2}Q_{ij}Q_{ji} + \frac{B}{3} Q_{ij}Q_{jk}Q_{ki} + \frac{C}{4} \left(Q_{ij}Q_{ji}\right)^2 \right.\\
    & + \frac{1}{2} L_1 (\partial_k Q_{ij})(\partial_k Q_{ji}) + \frac{1}{2} L_2 (\partial_i Q_{jk})(\partial_j Q_{ik}) \\ 
    & + \left. \frac{1}{2} L_3 Q_{ij} (\partial_i Q_{kl})(\partial_j Q_{kl}) + 2L_{1}q_0 \epsilon_{ikl}Q_{ij}(\partial_k Q_{lj}) \right) \mathrm{d}V.
\end{align*}
Here, $A$, $B$ and $C$ are material parameters and $L_1$, $L_2$, and $L_3$ are elastic constants in the tensorial formulation of the elastic free energy,
and $q_0=2\pi/P_0$ is the inverse chiral pitch. $L_1$, $L_2$, and $L_3$ can be computed from the elastic constants of the
splay $K_1$, twist $K_2$, and bend $K_3$ deformation modes that are formulated within the director-based Frank free energy.
Often, single elastic constant approximation $(L_1 = L$,  $L_2=L_3=0)$ is used, where splay, twist and bend elastic modes have equal contributions ($K_1=K_2=K_3=K$). In Figs.~\ref{fig:elastic_constants_L21} and \ref{fig:elastic_constants_K123}, we explore the role of elastic anistropy, i.e. how individual elastic modes influence the active nematic dynamics and we use non-zero $L_1, L_2 \text{ and } L_3$ elastic constants, following the relations
\begin{align}
    L_1 &= \frac{2}{27 S^2} (-K_1 + 3K_2 + K_3),\nonumber\\
    L_2 &= \frac{4}{9 S^2} (K_1 - K_2),\label{eq:mapping}\\
    L_3 &= \frac{4}{27 S^3} (-K_1 + K_3).\nonumber
\end{align}

In the formulation of the elastic constants in Eqs. \ref{eq:mapping}, $K_{24}$ is equal to $\frac{K_1}{2}$ but is not relevant due to periodic boundary conditions.
The flow field obeys the incompressibility condition and the Navier-Stokes equation,
\begin{align}
\partial_i  v_i &= 0,   \label{eq:continuity}\\ 
\rho (\partial_t + v_j \partial_j) v_i&= \partial_j \Pi_{ij},\label{eq:navier-stokes}    
\end{align}
where $\rho$ is the fluid density and $\Pi_{ij}$ is the stress tensor, which consists a passive and an active term $\Pi_{ij} = \Pi_{ij}^\text{passive} + \Pi_{ij}^\text{active}$, where
\begin{align}
       \Pi_{ij}^\text{passive} = &-p \delta_{ij} + 2 \eta D_{ij} + 2\chi \left( Q_{ij} +\frac{\delta_{ij}}{3} \right) Q_{kl} H_{kl} \\
       &- \chi H_{ik} \left( Q_{kj} +\frac{\delta_{kj}}{3} \right) - \chi \left( Q_{ik} +\frac{\delta_{ik}}{3} \right)H_{kj}\\
       &+ Q_{ik} H_{kj} - H_{ik} Q_{kj} - \partial_i Q_{kl}\frac{\delta F}{\delta \partial_j Q_{kl}},\label{eq:passive}\\
       \Pi_{ij}^\text{active} &= -\zeta Q_{ij}.
\end{align}
Here, $p$ is the fluid pressure, $\eta$ is the isotropic viscosity, $\chi$ is the flow alignment parameter and $\zeta$ is the activity, which is positive in extensile materials and negative in contractile materials.

The coupled equations for the nematic order $Q_{ij}$ and fluid velocity $v_{i}$ are numerically solved using the hybrid lattice-Boltzmann approach \cite{CarenzaLN_ProcNatlAcadSci116_2019, ZhangR_NatCommun7_2016,thampi2014vorticity,kralj2023defect,KrajnikZ_SoftMatter16_2020,CoparS_PhysRevX9_2019}, based on the finite difference method for solving the Q-tensor evolution (Eq. \ref{eq:q_dynamics}), and the D3Q19 lattice Boltzmann method for the incompressibility and the Navier-Stokes equation (Eqs. \ref{eq:continuity}, \ref{eq:navier-stokes}).

\subsection*{Material parameters}

In the paper, we consider a single elastic constant approximation for the material constants ($L_1 \neq 0$, $L_2=0$, and $L_3=0$), except in Figs.~\ref{fig:elastic_constants_L21} and \ref{fig:elastic_constants_K123}. Changing the values of the elastic constants affects the nematic correlation length and in turn the resolution of the numerical mesh resolution. Accordingly, we 
compute the average Frank elastic constant
$$\overline{K} = \frac{1}{3}(K_1 + K_2 + K_3)=\frac{9S^2}{2}\left( L_1 + \frac{L_2}{3} \right)$$
from the mapping in Eqs.~\ref{eq:mapping}.
To account for simulation results at different values of $\overline{K}$ in Fig.~\ref{fig:elastic_constants_L21}, we set $L_1=L$, vary the elastic constant $L_2$, and express the simulation results with a constant term $K_0 = \frac{9S_\text{eq}^2}{2}L$. Here, $L$ represents the fixed value of the elastic constant as used in Figs.~\ref{fig:alignment} and \ref{fig:q0_rho_error}.
In Figure 4, where we explore the role of the elastic anisotropy, we use the condition of a constant average Frank elastic constant $\overline{K}$ and choose the Frank elastic constants accordingly and express the simulation results in terms of $K_0$. From a given set of the Frank elastic constants, the tensorial elastic constants $L_1$, $L_2$, and $L_3$ are determined from a mapping given by Eqs.~\ref{eq:mapping} evaluated at $S=S_\text{eq}$.

Using the alignment parameter $\chi$, rotational viscosity parameter $\Gamma$ and isotropic viscosity $\eta$ from the Beris-Edwards model of nematodynamics (Eqs.~\ref{eq:q_dynamics},  \ref{eq:passive}), we can express the Ericksen-Leslie viscosity parameters~\cite{deGennesPG_1993, denniston2001lattice} that are typically formulated within the director-based approach to nematodynamics:
\begin{align*}
\alpha_1&=\frac{\chi^2}{\Gamma}\frac{9S^2}{2}\left(3S^2-2S-1\right),\\
\alpha_2&=-\frac{\chi}{\Gamma}\frac{S}{4}\left(3S+4\right)-\frac{1}{\Gamma}\frac{9S^2}{4},\\
\alpha_3&=-\frac{\chi}{\Gamma}\frac{S}{4}\left(3S+4\right)+\frac{1}{\Gamma}\frac{9S^2}{4},\\
\alpha_4&=\frac{\chi^2}{\Gamma}\left(S-\frac{2}{3}\right)^2+2\eta,\\
\alpha_5&=-\frac{\chi^2}{\Gamma}\frac{S}{4}\left( 3S-8\right)+\frac{\chi}{\Gamma}\frac{S}{4}\left( 3S+4\right),\\
\alpha_6&=-\frac{\chi^2}{\Gamma}\frac{S}{4}\left( 3S-8\right)-\frac{\chi}{\Gamma}\frac{S}{4}\left( 3S+4\right).
\end{align*}
The flow-alignment parameter $\lambda$ in the director-based formulation then reads as $\lambda=\frac{\alpha_2+\alpha_3}{\alpha_2-\alpha_3}=\frac{3S+4}{9S}\chi$ and the flow-aligning to flow-tumbling transition occurs at $\lambda=1$~\cite{deGennesPG_1993}.

The overall results of the simulations are expressed in the units of the elastic constant $L$, rotational viscosity parameter $\Gamma$ and mesh resolution $\Delta x$. Mesh resolution is defined as $\Delta x = 1.5\chi_\text{n}$, where $\chi_\text{n}$ is nematic correlation length defined as $\chi_\text{n} = \sqrt{L/(A+BS_{\text{eq}} +\frac{9}{2} CS_{\text{eq}}^2)}$, where $S_{\text{eq}}=0.533$ is equilibrium degree of nematic order, $A = -0.43 L/(\Delta x)^2$, $B=-5.3L/(\Delta x)^2$, $C=4.325L/(\Delta x)^2$ and $\eta = 1.38/\Gamma$. The size of the simulation box is $201\times201\times201$ mesh points and periodic boundary conditions are used in all three spatial directions. $200\times200\times200$ simulation box is used in Fig.~\ref{fig:alignment}. The time step in simulations is set to $\Delta t = 0.025 (\Delta x)^2/(L\Gamma)$.
To recover the simulation results in the SI units, one can use typical parameter values for active systems $L=3\,\text{pN}$, $\Delta x=2\,\upmu\text{m}$, and $\Gamma=10\,\text{(Pa s)}^{-1}$, roughly estimated from active turbulence in bacterial and microtubule systems~\cite{WolgemuthCW_BiophysJ95_2008, SanchezT_Nature491_2012} and viscoelastic properties of lyotropic nematics~\cite{zhang2018interplay}.

\subsection*{Data analysis}
We define the defect density as the length of defect lines over a unit volume and compute it from the defect volume fraction of the regions where scalar order parameter is $S<0.4$~\cite{kralj2023defect}. In the next analysis step, we average either the defect density (Figs.~\ref{fig:alignment} and \ref{fig:elastic_constants_K123}) or the inverse defect density (Figs.~\ref{fig:elastic_constants_L21} and \ref{fig:q0_rho_error}) in time and calculate its standard deviation that is presented with \emph{error} bars. For the velocity field, we compute the average of the velocity squared $\langle v^2\rangle _V$ over the complete simulation volume $V$ at given time. In Figs.~\ref{fig:alignment}, \ref{fig:elastic_constants_K123}, and \ref{fig:q0_rho_error}, we obtain the root mean square velocity as $\langle v \rangle =\sqrt{\langle v^2\rangle _V}$ and the variability (presented by error bars) from its standard deviation. In Fig.~\ref{fig:elastic_constants_L21}, the mean inverse velocity squared $\langle 1/v^2 \rangle$ and its standard deviation are obtained from the time-dependence of the  $1/\langle v^2\rangle _V$. 
To estimate the value of the critical pitch for the transition from the active chiral turbulence to the active blue phase, we performed nonlinear regression on simulation data for mean velocity and mean inverse defect density in Fig. 5. The value of $P_\text{c}$ with the precision of $1\,\Delta x$ was determined so that the mean velocity in the log-log plot (Fig. ~\ref{fig:q0_rho_error}(i)) shows a power-law scaling for the longest range of the pitch values. Once $P_\text{c}$ was set, we employed a similar procedure to determine $1/\rho_\text{c}$ based on the log-log plot of the mean inverse defect density (Fig.~\ref{fig:q0_rho_error}(g)).

\vspace{1cm}

\subsection*{Data availability}
The data sets generated in this study are available from the corresponding author upon reasonable request.

\subsection*{Code availability}
The code used in this study is available from the corresponding author upon reasonable request.

\bibliography{collection}

\begin{acknowledgments}
The authors acknowledge funding from Slovenian Research and Innovation Agency (ARIS) under contracts P1-0099, J1-50006, J1-2462, N1-0195. This result is also part of a project that has received funding from the European Research Council (ERC) under the
European Union’s Horizon 2020 Research and Innovation Program (Grant Agreement No. 884928-LOGOS).
\end{acknowledgments}

\vspace{1cm}

\subsection*{Author contributions}
N.K. performed the numerical simulations and the analysis. M.R. and Ž.K. designed and supervised the research. All authors contributed to the writing of the manuscript.

\subsection*{Competing interests}
The authors declare no competing interests.

\end{document}